\documentclass[aps,twocolumn,longbibliography,groupedaddress]{revtex4-2}%

\usepackage{amsmath}
\usepackage{amsfonts}
\usepackage{amssymb}
\usepackage{graphicx}%
\providecommand{\U}[1]{\protect\rule{.1in}{.1in}}
\begin{document}

\title{Notes on altermagnetism and superconductivity}
\author{Igor I. Mazin}
\affiliation{Department of Physics \& Astronomy, George Mason University, Fairfax, VA 22030, USA}
\affiliation{ Quantum Science and Engineering
Center, George Mason University, Fairfax, VA 22030, USA.}
\maketitle

Altermagnetism is a recently discovered new type of collinear magnets, which
share some characteristic features with ferromagnets (lack of the
nonrelativistic Kramers degeneracy at a general point in the Brillouin zone,
finite anomalous Hall effect, finite magnetooptical effect) and other with
antiferromagnets (net magnetization zero by symmetry)\cite{AM}. While numerous
properties of altermagnets have been explored, largely from the point of view
of spintronics, interplay between superconductivity and altermagnetism, another aspect in which ferro- and antiferromagnets are
principally different, has not been addressed so far. Not surprisingly, there altermagnets can can manifest
properties typical for ferromagnets in one contexts, and those typical for
antiferromagnets in another. 

There are two issues that are typically considered in terms of interaction
between magnetism and superconductivity:  (1) what kind of superconducting
state may be consistent with a given magnetic order and (2) what kind of
pairing can be generated by proximity to a magnetic order (in other words, if
we can gradually suppress the long range magnetic order by an external
stimulus, such as pressure, what supperconducting symmetry may emerge on the
either side of the quantum crutical point?).

\subparagraph{Superconductivity and ordered altermagnetism}

It is well known that the standard antiferromagnetism can support singlet
superconductivity (as for instance in Fe-based superconductors) as long as the
coherence length is much larger than the period of antiferromagnetic order. On
the other hand, a split ferromagnet with spin-split bands ($i.e.$, the eigenvalues $\epsilon
_{\mathbf{k\uparrow}}\neq\epsilon_{-\mathbf{k\downarrow}}$) can only support
Cooper pairs with the spinor order parameter $\Delta_{\uparrow\uparrow},$
which is triplet. A standard representation\cite{SU}  of the spinor triplet
order parameter in terms of a spacial vector $\mathbf{d}$ describe this spinor
as
\begin{equation}
\Delta_{\alpha\beta}=%
\begin{pmatrix}
-d_{x}+id_{y} & d_{z}\\
d_{z} & d_{x}+id_{y}%
\end{pmatrix}
\end{equation}
Obviously, since only the $\alpha=\beta=\uparrow$ element is nonzero for a
given $\mathbf{k}$, $d_{z}=0,$ and $d_{x}=-id_{y}.$

AM is rather close to FM in this sense, but it has an additional symmetry:
there is an element of the point group that does not map the Fermi surface for
a given spin upon itself, but does map it upon the Fermi surface for the
opposite spin\cite{AM,PNAS}. Let us now, for simplicity, consider a tetragonal
FM material. Then the only triplet state consistent with this requirement is
the nonunitary state \textbf{d}$=F_{1}(k)k_{z}(\hat{x}+i\hat{y}),$ where $F$
has the full tetragonal symmetry, so that of $\Delta_{\alpha\beta}$ only
$\Delta_{\uparrow\uparrow}\neq0$. If we now consider the other Fermi surface, for the
opposite spin, the order parameter there will be $\mathbf{d=}F_{2}%
(k)k_{z}(\hat{x}-i\hat{y}),$ with only $\Delta_{\downarrow\downarrow}\neq0.$
These two order parameters have different symmetries, and are not degenerate,
so the critical temperature $T_{c}$ will be different. The symmetry
consideration dictate that, without spin-orbit interaction, there will be no
coupling between the two order parameters. In a typical experiment probing the
average order parameter such a system at low temperature will behave as mixed
state $\mathbf{d=}\frac{k_{z}}{2}\mathbf{[}F_{1}(k)+F_{2}(k)]\hat{x}%
+i\frac{k_{z}}{2}\mathbf{[}F_{1}(k)-F_{2}(k)]\hat{y}.$ This state is
\textit{nonunitary} as long as $F_{1}\neq F_{2},$ and nematic (breaks the
$C_{4}$ symmetry).

AM, despite the absence of the net magnetization, behaves very much like a
ferromagnet in the sense that any Cooper paper can be either $\Delta
_{\uparrow\uparrow}$ or $\Delta_{\downarrow\downarrow}.$ Correspondingly, the
order parameters will be $F_{1}(k)k_{z}(\hat{x}+i\hat{y})$ and $F_{2}%
(k)k_{z}(\hat{x}-i\hat{y}).$ However, in this case $F_{1}=F_{2}$ by symmetry
(the same symmetry that transfoms one spin sublattice into the other),
so the average order parameter will be just $\mathbf{d=}k_{z}F(k)\hat{x}.$
This order parameter is strictly \textit{unitary} (correspondingly, the
condensate is not spin-polarized, just as the normal state isn't), and
nematic. Of course, the partner state $\mathbf{d=}k_{z}F(k)\hat{y}$ will be
degenerate with this one. In this sense, the AM as regards superconductivity
again has some features similar to ferromagnets, some similar to
antiferromagnets, and some unique. An interesting analogy may be drawn with
the Ising superconductivity, appearing when the Kramers degeneracy is lifted
not by the exchange field, but by the spin-orbit coupling. In that case the
two spin-split Fermi surfaces carry order parameters that are strictly
$S+T$ and $S-T,$ where $S(T)$ stands for singlet(triplet)\cite{PRX}. Despite
that, in most experiments, namely those that probe the average order
parameters, they behave \textit{approximately} as singlet (approximately
because no symmetry requires the two order parameters to be exactly the same).
In case of AM the difference is that the average order parameter becomes
unitary \textit{exactly}, by symmetry (of course, remaining triplet)

\subparagraph{Superconductivity and altermagnetic fluctuations}

A related question is what superconducting symmetry can be generated by the
AM-type spin fluctuations. Before discussing that let us compare FM and AF
fluctuations a bit more carefully than how it is usually done. 

In case of ferromagnetic fluctuations, the spin fluctuation spectrum is peaked
at $\mathbf{q}=0.$ Given that spin fluctuations are repulsive in the singlet
channel (the partners in a Cooper pair interact to a spin fluctuations with
opposite signs), and that by continuity $\Delta(\mathbf{k)\approx}%
\Delta(\mathbf{k+q),}$ as long as $\mathbf{q}$ is small, such fluctuations
will always be pair-breaking. Traditional, N\'{e}el type AF order at a finite
vector $\mathbf{q}$ (if $\mathbf{q}$ lies at the zone boundary, this order
will correspond to doubling of the unit cell). In that case spin fluctuations
can be pairing as long as  $\Delta(\mathbf{k)\cdot}\Delta(\mathbf{k+q)}<0.$
Popular theories ascribing the $d$-wave superconductivity in cuprates
($\mathbf{q}_{2D}=\{\pi,\pi\})$ and the $s_{\pm}$ superconductivity in Fe
pnictides ($\mathbf{q}_{2D}=\{\pi,0\})$ to spin fluctuations utilize this
property. While such fluctuations can also generate a $p$-wave pairing,
especially when combined with an anisotropic electron-phonon
coupling\cite{Amy}, in practice it is very difficult\cite{Markus}. 

In such discussions it is routinely assumed that AF fluctuations
\textit{always} correspond to a finite $\mathbf{q.}$ However, several hundreds
of known antiferromagnets have magnetic order corresponding to $\mathbf{q=0.}$
This can happen, of course, if the magnetic species occupies a Wyckoff
position with a multiplicity larger that one. For instance, the very popular
now family of Kagome superconductors, $A$V$_{3}$Sb$_{5}$ ($A$ is an alkaline
metal) is believed to host spin-fluctuations at $\mathbf{q\approx0}$ and the
intra-triangular correlations of 120$%
{{}^\circ}%
.$ Of course, magnetic order and spin fluctuations  $\mathbf{q\approx0}$ and
collinear spins are also perfectly possible, and altermagnetic (and some conventional
antiferromagnets) belong to this class.

\begin{figure}[ht]
\includegraphics[width=0.45\linewidth]{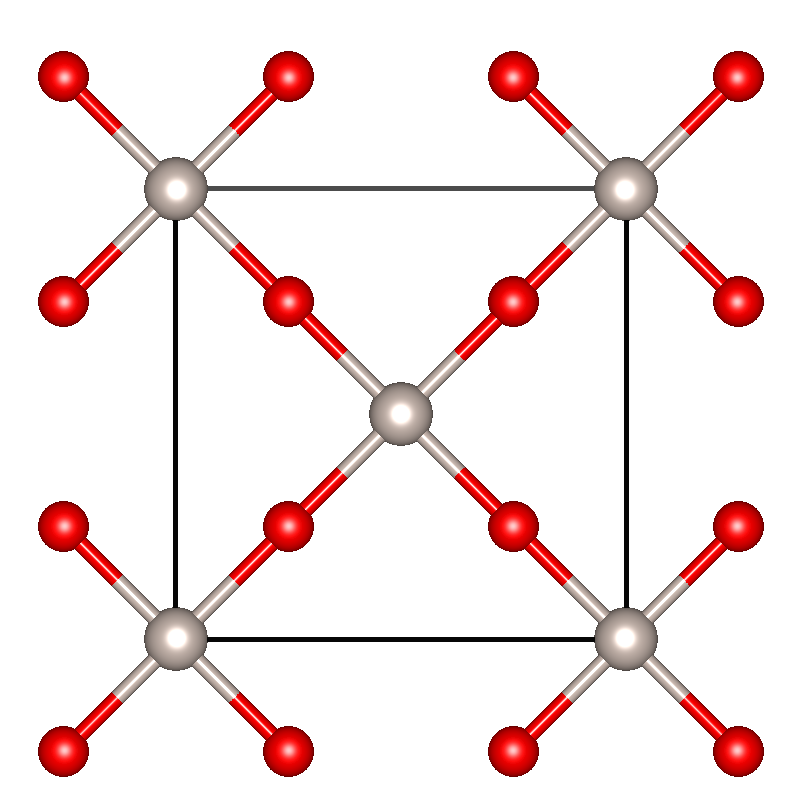}
\includegraphics[width=0.45\linewidth]{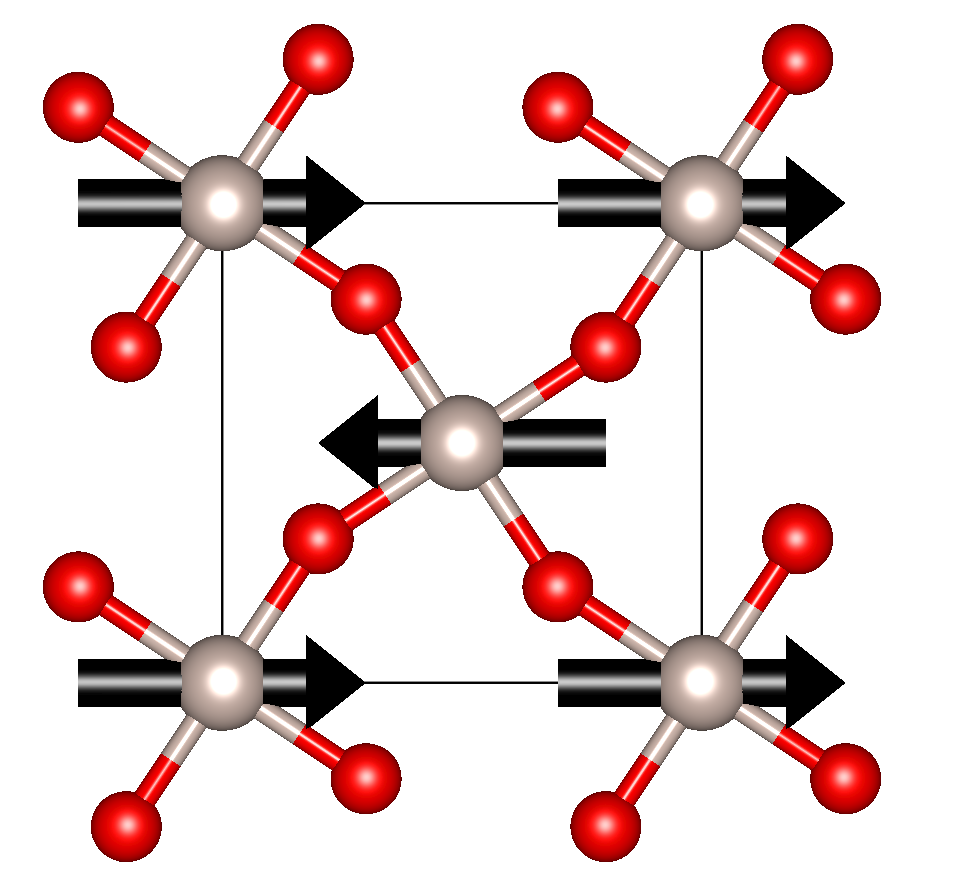}
\caption{(left) An example of a 2D P4/mmm structure. The grey balls are metallic (M), and potentially magnetic
ions, and the red ones are ligands (L). (right) Same for a P4/mbm structure, which can carry 
an altermagnetic state (shown by arrows)}%
\label{123}%
\end{figure}

\begin{figure}[ht]
\includegraphics[width=0.45\linewidth]{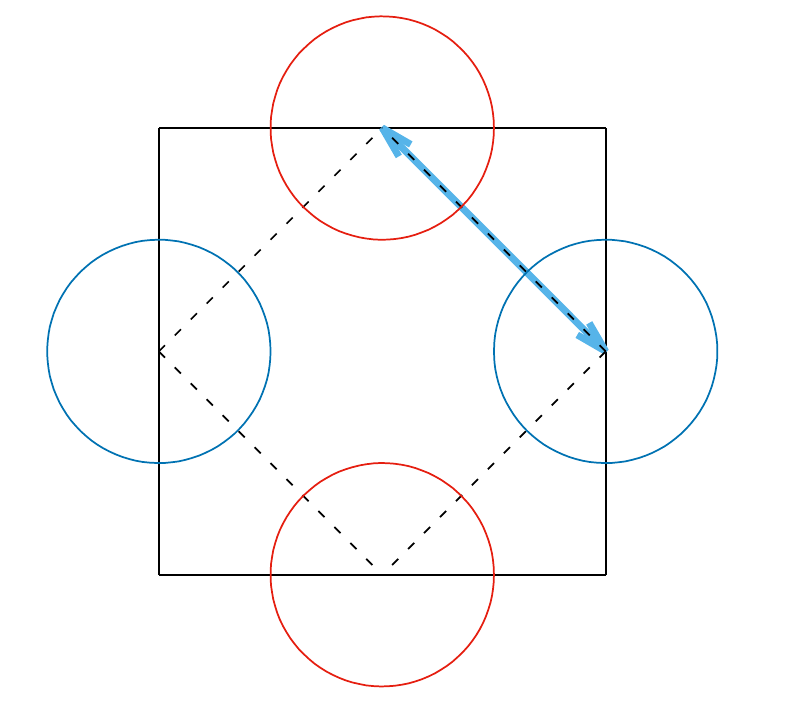}
\includegraphics[width=0.45\linewidth]{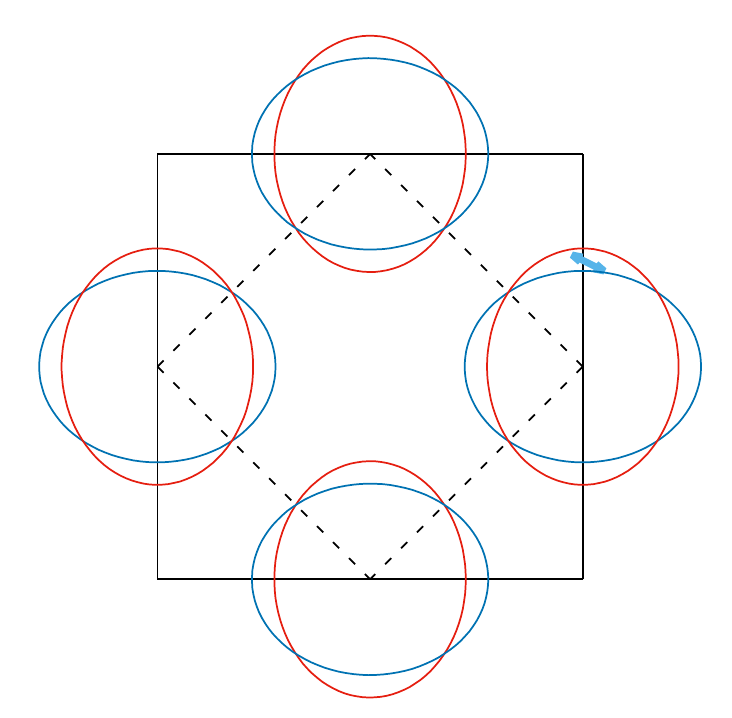}
\caption{(left) A possible 2D Brillouin zone consistent with the crystal symmetry shown in Fig. 
\protect\ref{123}(left). (right) The same, after downfolding corresponding to the double 
unit cell shown in Fig. \protect\ref{123}(right). The downfolded zone is shown with the dashed lines. Colors 
reflect the signs of a possible $d$-wave (in the unfolded zone) order parameter. The arrows give examples of
antiferromagnetic spin fluctuations with $\mathbf{q}\approx {\pi,\pi}$ generating a 
$d$-wave pairing. Nothe that the case on the right these are fromally  altermagnetic fluctuations.}%
\label{BZ}%
\end{figure}

In order to understand the physics of  $\mathbf{q\approx0}$ spin fluctuations
in the context of superconductivity, let us consider a hypothetical 2D lattice
depicted in Fig. \ref{123}(left). Here M is a metal ion and L is a ligand. This is a
tetragonal structure with the symmetry group I4/mmm. Lets us assume that this
structure generates a Fermi surface centered around the X(Y) points, as shown
in Fig. \ref{BZ} (left), and spin-fluctuations corresponding to $\mathbf{q=}\{\pi
,\pi\}.$This model was introduced by Agterberg $et$ $al$\cite{ABG}, and it
leads a $d$-wave superconductivity of the type $k_{x}^{2}-k_{y}^{2}.$ 

Let us now introduce a small distortion, rotations of the $ML_{2}$ squares,
shown in Fig. \ref{123}(right). The symmetry group is now P4/mbm, it is also
tetragonal, but has now two metal ions per cell. This will lead to folding
down of the original Brillouin zone (Fig.  \ref{BZ}), so that now there are
to Fermi contours around each M point; if the original Fermi contours were
close to circular, the downfolded one will be nearly degenerate.

Altermagnetic spin fluctuations will have $\mathbf{q\approx0},$ however, this
does not mean that, as in for ferromagnetic fluctuations, such fluctuations
can only generate triplet pairing. In fact, since in this particular example
the AM order only slightly deviates from the AF order, the generated pairing
state must be close to the downfolded $d$-state of Ref. \cite{ABG}. In
principle, the two crossing Fermi lines will hybridize, and the order
parameter form nodal lines, as discussed in Ref. \cite{KFS}. However, this is a
relatively unimporant effect. 

From the formal point of view, the issue is that $\mathbf{q\approx0}$ the spin
susceptibility may have importand internal structure, and has to be written as
$\chi(\mathbf{q,r}_{1},\mathbf{r}_{2}),$ where $\mathbf{r}$ is defined inside
the first unit cell, or as a matrix in reciprocal vectors, $\chi
(\mathbf{q+G,q+G}^{\prime}).$ The corresponding vertex will be determined by
the variation of the (nonmagnetic) one-electron Green function with respect to
a fluctuation generating opposite magnetic moments on the two sublattices. A
more detailed theory than that usually used for ferro- or antiferromagnetic
spin fluctuations needs to be developed, and it  is not unreasonable to
assume that both triplet and singlet pairing can be iduced by AM spin
fluctuations, depending on the details.

Lastly, one can make another interesting observation: so far, two very
different classes of superconductors offer protection from thermodynamic pair
breaking (Pauli limiting) for some directions of magnetic fields. The first
one are triplet superconductors without net magnetization (such as $^{3}$He or
the initial (now debunked) model for Sr$_{2}$RuO$_{4})$; there the spin
susceptibility in the superconducting state is the same as in the normal
state, $\chi_{sc}=\chi_{n}$, due to triplet pairs having the same ability to
screen the field (in some directions) as the individual electrons. Unitary
triplet superconductivity, often discussed in connection with some
ferromagnetic U compounds, is also protected, and again $\chi_{sc}=\chi_{n},$
but there the underlying mechanism is different: If there is and easy-axis
magnetocrystalline anisotropy in the normal state, then screening of a small
external field is afforder not by increasing the net number of electrons in
one spin channel at the expense of the other, but by canting spins of
electrons removed from the Fermi level. Indeed, if an external field $H$ is
applied perpendicular to this axis, the Fermi surface does not change in the
linear in $H$ order, but a linear in $H$ magnetization does appear, on the
order of $H\Delta\Omega/\left\langle \delta V_{xc}\right\rangle ,$ where
$\Delta\Omega$ is the volume difference between the two spin-split Fermi
surfaces and $\left\langle \delta V_{xc}\right\rangle $ is the properly
averaged exchange splitting. Obviously, the ratio $\Delta\Omega/\left\langle
\delta V_{xc}\right\rangle $ is a number on the order of the density of
states, so this provides a contribution to susceptibilty on the order of the
Pauli susceptibility, and is not affected by opening a superconducting gap
$\Delta\ll\left\langle \delta V_{xc}\right\rangle .$ Note that similar
protections is operative in Ising superconductors, where the role of $\delta
V_{xc}$ is played by the spin-orbit coupling\cite{PRX}. In the spirit of the
key feature of AM, namely sharing features of both FM and AFM, they also have
a Pauli protection, but which in this case is similar to that in ferromagnets,
so it does not require accounting for relativistic effects in the band
structure, but requires a magnetic anisotropy. 

In the above discussion, we have addressed issues related to possible
coexistance of AM and superconductivity, as well as superconductivity possibly
induced by AM spin fluctuation. A further step in investigating the interplay
between magnetism and superconductivity would involve possible effects at the
\textit{interface} between a conventional superconductor and an altermagnet. 

One of the most interesting effects in this environment is spin-polarized
Andreev reflection. Andreev reflection at a boundary between a conventional
superconductor and a ferromagnet is well understood\cite{PCAR} and is often
use to measure the transport spin polarization of ferromagnets. In a generic
ferromagnet, as opposed to a traditional antiferromagnet, the number of
conductivity channels for two spins are not the same (in other words, the area
of the Fermi surface projection onto the interface is spin-dependent. Since an
Andreev process consist of a spin-up electron with a momentum $\mathbf{k}$ and
a spin-down one with a momentum $-\mathbf{k,}$ some electrons will never find
a partner and therefore the conventional Andreev conductivity, which is twice
the normal conductivity, will be suppressed. So defined spin polarization
depends on the orientation of the interface. While for a ferromagnet is can
only be zero by accident, in an AM, for particular interface orientations the
number of conductivity channels is the same for both spins, therefore one
expects no suppression, just as in an AF, but in some other, and in fact in
general directions the areas of the two projections will be different, and a
finite suppression will be measured. Again, an AM sometimes behaves as an AF,
and sometimes as a FM.
\begin{figure}[ht]
\includegraphics[width=0.45\linewidth]{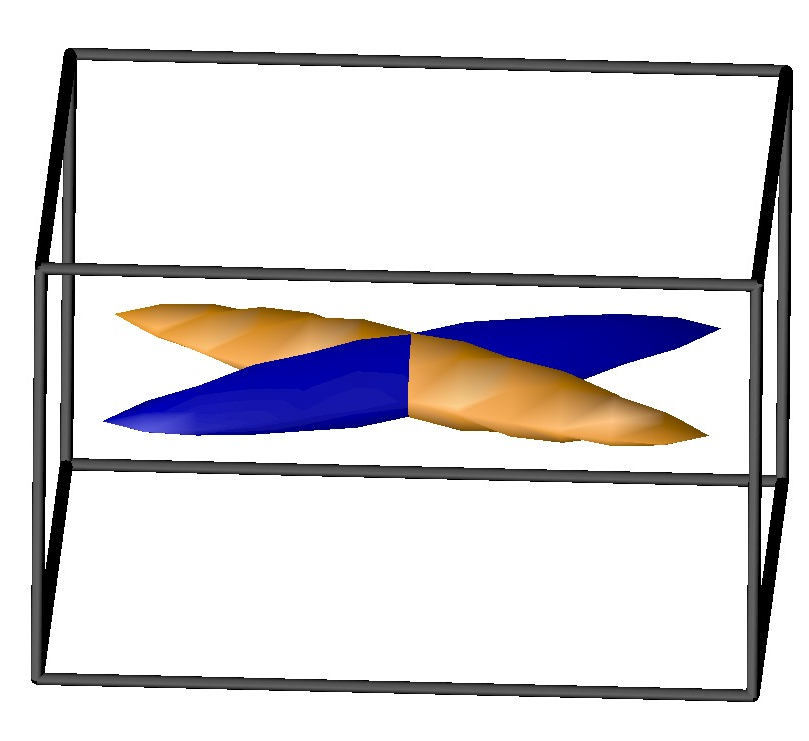}
\includegraphics[width=0.45\linewidth]{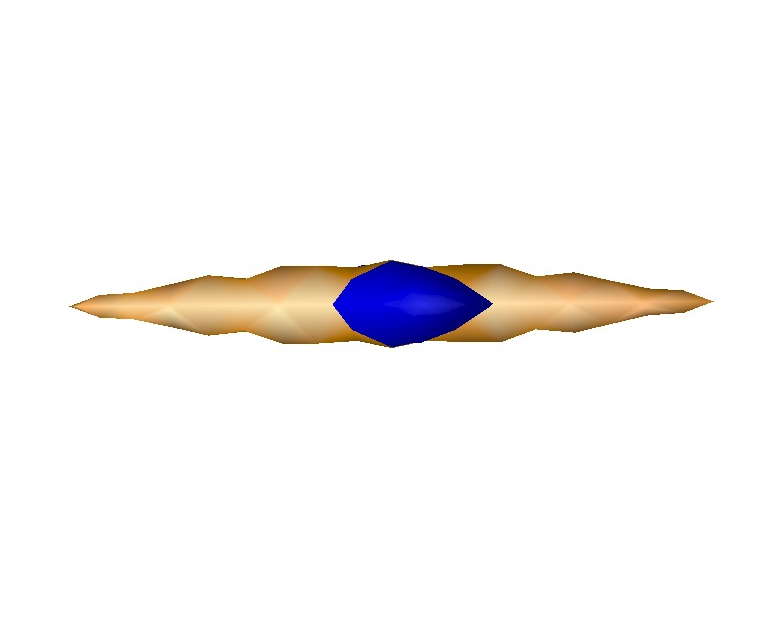}
\caption{(left) A possible 2D Brillouin zone consistent with the crystal symmetry shown in Fig.
\protect\ref{123}(left). Specifically, a cut-off of a specific Fermi surface pocket in the hypothetical
altermagnetic FeSb$_2$\protect\cite{PNAS}. Different colors denote different spins. (right) Projection of this pocket onto the (110) inteface.}%
\label{And}%
\end{figure} 
This is illustrated in Fig. \ref{And}(left), where a single pocket of the Fermi
surface of the hypothetical AM FeSb$_{2}$\cite{PNAS} at a particular Fermi
energy is cut off to show the symmetry. Evidently, for a \{100), or (010), or
(001) interface every  $\mathbf{k}_{||}$ has a partner with $-$ $\mathbf{k}%
_{||}$ and the opposite spin, for the (110) interface, for instance, this is
not the case, as is quite obvious from Fig.\ref{And}(right), where the projections of
the two Fermi surfaces are shown. 

In this note we have analyized various aspects of interplay between the novel
magnetic phenomenon, altermagnetism, and superconductivity. This analysis
should be helpful in designing new experiments to further study this unusual state.

Helpful discussions with Maxim Khodas and financial support from DOE under the grant DE-SC0021089 are acknowledged.


\begin{thebibliography}{99}                                                                                               %
\bibitem {AM}Libor \^{S}mejkal, Jairo Sinova, Tomas Jungwirth, Altermagnetism:
spin-momentum locked phase protected by non-relativistic symmetries,
arXiv:2105.05820 (20121)

\bibitem {SU}Manfred Sigrist and Kazuo Ueda, Phenomenological theory of
unconventional superconductivity, Rev. Mod. Phys. \textbf{63}, 239 (1991)

\bibitem {Amy}I. Schnell, I. I. Mazin, A.Y. Liu, Unconventional
superconducting pairing symmetry induced by phonons, Phys. Rev. \textbf{B 74},
184503 (2006)

\bibitem {Markus}P. Steffens, Y. Sidis, J. Kulda, Z. Q. Mao, Y. Maeno, I.I.
Mazin, and M. Braden, Spin fluctuations in Sr$_{2}$RuO$_{4}$ from polarized
neutron scattering: implications for superconductivity. Phys. Rev. Lett.
\textbf{122}, 047004 (2019).

\bibitem {125}Brenden R. Ortiz, L\'{\i}dia C. Gomes, Jennifer R. Morey, Michal
Winiarski, Mitchell Bordelon, John S. Mangum, Iain W. H. Oswald, Jose A.
Rodriguez-Rivera, James R. Neilson, Stephen D. Wilson, Elif Ertekin, Tyrel M.
McQueen, and Eric S. Toberer, New kagome prototype materials: discovery of
KV$_{3}$Sb$_{5}$, RbV$_{3}$Sb$_{5}$, and CsV$_{3}$Sb$_{5},$ Phys. Rev.
Materials\textbf{ 3}, 094407 (2019)

\bibitem {PNAS} Igor I. Mazin, Klaus Koepernik, Michelle D. Johannes, Rafael
Gonz\'{a}lez-Hern\'{a}ndez, and Libor \v{S}mejkal, Prediction of
unconventional magnetism in doped FeSb$_{2},$ PNAS \textbf{118,} e2108924118 (2021)

\bibitem {PRX}D. Wickramaratne, S. Khmelevskyi, D. F. Agterberg, and I. I.
Mazin, Ising superconductivity and magnetism in NbSe$_{2}$, Phys. Rev. X
\textbf{10}, 041003 (2020)

\bibitem {ABG}D. F. Agterberg, Victor Barzykin, and Lev P. Gor'kov,
Conventional mechanisms for exotic superconductivity.Phys. Rev. B \textbf{60},
14868 (1999)

\bibitem {KFS}I.I. Mazin, Symmetry analysis of possible superconducting states
in K$_{x}$Fe$_{2}$Se$_{2}$ superconductors, Phys. Rev. B \textbf{84}, 024529 (2011)

\bibitem {PCAR}I.I. Mazin, A.A. Golubov, and B. Nadgorny, Probing Spin
Polarization with Andreev Reflection: A Theoretical Basis. J. Appl. Phys.,
\textbf{89}, 7576 (2001);  G.T. Woods, R, J. Soulen Jr., I. I. Mazin, B.
Nadgorny, M. S. Osofsky, J. Sanders, H. Srikanth, W. F. Egelhoff and R. Datla,
Analysis of Point-contact Andreev Reflection Spectra in Spin Polarization
Measurements. Phys. Rev. B \textbf{70}, 054416 (2004).
\end{thebibliography}
\end{document}